\documentclass{aa501}

\usepackage{graphicx}
\usepackage{natbib}
\bibpunct{(}{)}{;}{a}{}{,} 

\newcommand{\La}{\mbox{${\rm Ly\alpha}$}}

\newcommand{\Rns}{\mbox{$R_{\rm ns}$}}
\newcommand{\Mns}{\mbox{$M_{\rm ns}$}}
\newcommand{\Tns}{\mbox{$T_{\rm ns}$}}

\newcommand{\Msun}{\mbox{$M_{\odot}$}}

\newcommand{\gc}{\mbox{$\rm g\;cm^{-2}$}}

\newcommand{\taumin}{\mbox{$\tau_{\rm min}$}}
\newcommand{\taumax}{\mbox{$\tau_{\rm max}$}}
\newcommand{\mint}[2]{\int\limits_{#1}^{#2}}

\newcommand{\D}{\displaystyle}

\begin{document}

\title{Thermal emission from low-field neutron stars}

\author{B.T. G\"ansicke\inst{1,2} \and
        T.M. Braje\inst{2} \and
        R.W. Romani\inst{2}}
\offprints{B. G\"ansicke, boris@uni-sw.gwdg.de}

   \institute{  Universit\"ats-Sternwarte G\"ottingen,
                Geismarlandstr. 11, D-37083 G\"ottingen, Germany
        \and    Physics Department, Stanford University, Stanford, 
                CA 94305-4060, USA
             }
\date{Received 23 July 2001 / Accepted 25 February 2002}

\abstract{We present a new grid of LTE model atmospheres for weakly
magnetic ($B\la10^{10}$\,G) neutron stars, using opacity and equation
of state data from the OPAL project and employing a fully frequency-
and angle-dependent radiation transfer. We discuss the differences 
from earlier models, including a comparison with a detailed NLTE
calculation. 
We suggest heating of the outer layers of the neutron star
atmosphere as an explanation for the featureless X-ray spectra of
RX\,J1856.5--3754 and RX\,J0720.4--3125 recently observed with
\textit{Chandra} and \textit{XMM}.
 }

\maketitle

\section{Introduction}
Modern X-ray observatories detect the thermal surface emission of a
number of neutron stars. In analogy to the classic model atmosphere
analysis of normal stars, such observations permit the direct
measurement of fundamental neutron star properties, such as their
effective temperatures, atmospheric abundances and surface
gravities. The latter point is especially of great importance, as an
accurate measurement of the surface gravity of a neutron star is (with
the distance known) equivalent to a measurement of its mass/radius
ratio. The observational confirmation/rejection of the predicted
neutron star mass-radius relations is a fundamental test of our
understanding of the physics of matter above nuclear densities.

The first neutron star model atmospheres involving realistic opacities
were computed by \citet[][ henceforth R87]{romani87-1}, using atomic
data from the Los Alamos Opacity Library and employing a simple
angle-averaged radiation transfer. As a major result, R87 could show
that the thermal emission of neutron stars differs substantially from
a Planck spectrum. For low-metallicity (helium)
atmospheres, the emitted spectrum is harder than the corresponding
blackbody spectrum. The spectra emitted from high-metallicity
atmospheres (carbon, oxygen or iron) are closer to a
blackbody distribution, but show strong absorption structures in the
energy range observable with X-ray telescopes.

\citet[][ henceforth RR96]{rajagopal+romani96-1} computed hydrogen,
solar abundance, and iron model atmospheres for low-field neutron
stars, using improved opacity and equation of state data from the OPAL
project \citep{iglesias+roger96-1}, but employing the same
radiation transfer as R87. Contemporaneously, a similar set of
low-field atmospheres, partially based on the same atomic OPAL data,
but employing a more sophisticated radiation transfer, was presented
by \citet[][ henceforth ZPS96]{zavlinetal96-1}. The RR96 and ZPS96
models, broadly confirming the results of R87, were applied to X-ray
observations of presumably low-field neutron stars, i.e. millisecond
pulsars \citep[RR96,][]{zavlin+pavlov98-1}, isolated neutron stars
\citep{pavlovetal96-1}, and transiently accreting neutron
stars in quiescent LMXBs
\citep[e.g.,][]{rutledgeetal99-1,rutledgeetal00-1,rutledgeetal01-1}.

In the case of strongly magnetic neutron stars
($B\sim10^{12}-10^{14}$\,G), both the observation and the computation
of the thermal surface emission is significantly more difficult, as
non-thermal magnetospheric radiation dilutes the surface emission and
as little atomic data is readily available (for a review of
the properties of matter in strong magnetic fields, see
\citealt{lai01-1}). Magnetic models for hydrogen and iron atmospheres
were computed by \citet{pavlovetal95-1} and \citet{rajagopaletal97-1},
respectively.

The wealth of high-quality neutron star observations expected from
\textit{Chandra} and \textit{XMM} justifies the computation of modern
model atmospheres.  In this paper, we present new model atmosphere
grids for low-field neutron stars that overcome a number of
shortcomings and errors in the RR96 and ZPS96 calculations and that
are made  available to the community.

\section{The model atmospheres\label{s-model_grids}}
The new grid of neutron star atmospheres was computed
combining a fully frequency- and angle-dependent radiation transfer
code \citep{gaensickeetal95-1} and the OPAL opacities from RR96. The
employed radiation transfer is part of a model atmosphere code that
has previously been used for the computation of white dwarf
atmospheres which have been widely applied to observations of
cataclysmic variables and white dwarf/main-sequence binary stars.

The model atmospheres were created under the following assumptions.
(1) Plane-parallel geometry. The atmosphere of a neutron star has a
vertical extension of a few centimeters at most, compared to a radius
of $\sim10$\,km. Its curvature is hence completely negligible.
(2) Hydrostatic equilibrium. For the extreme surface gravity of
neutron stars and the relatively low temperatures that we consider
here, the atmosphere is static. The situation is different in
accreting neutron stars during X-ray bursts, where the nuclear
burning atmosphere considerably expands on short time scales.
(3) Radiative equilibrium. The atmosphere contains no source of energy,
but acts only as a blanket, through which the thermal energy of the
underlying core leaks out. In Sect.\,\ref{s-heating}, we will discuss
how abandoning this assumption will impact the spectrum of the neutron star.
(4) Local thermal equilibrium (LTE). For the high densities and the
low temperatures in the considered atmospheres collisional ion-ion
interactions dominate over interactions between matter and the
radiation field throughout most of the atmosphere. Deviations from LTE
in the outermost tenuous layers of the atmosphere may somewhat affect 
the depth of the absorption lines, but considering the overall level
of uncertainty in the  atomic physics involved, and the relatively poor
quality of the available spectroscopy, the assumption of LTE appears
to be a reasonable approximation. Below, we discuss a quantitative
comparison to a NLTE model.

With these approximations, the computation of a model atmosphere can
be separated into two independent parts: calculating the structure of
the atmosphere from the integration of the hydrostatic equation
(Sect.\,\ref{s-atmostruct}) and solving the radiation transfer
(Sect.\,\ref{s-radtrans}). The temperature structure of the
atmosphere, $T(z)$, is a free parameter in the overall process and is
adjusted iteratively until radiative and hydrostatic equilibrium are
satisfied (Sect.\,\ref{s-atmobuild}).

\begin{figure*}
\includegraphics[angle=270,width=18cm]{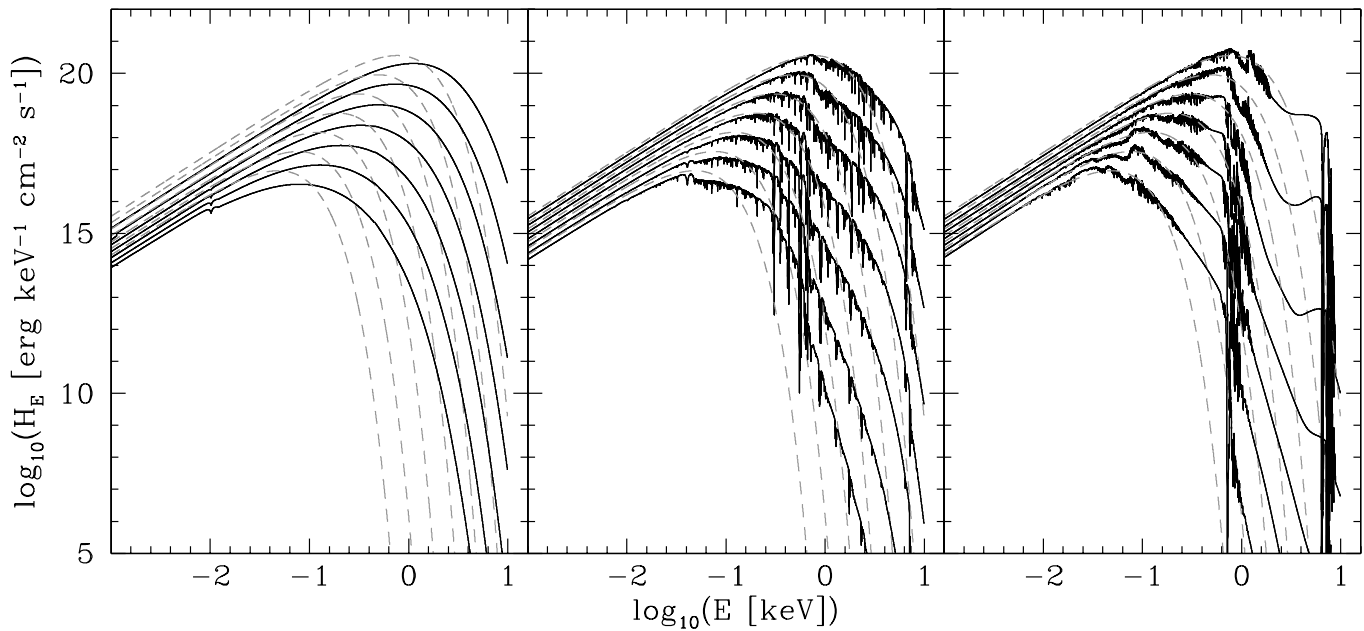}
\caption[]{\label{f-spectra} Emergent model spectra (unredshifted
Eddington flux) for $\Mns=1.4$\,\Msun, $\Rns=10$\,km neutron
stars. From left to right: hydrogen, solar abundance, and iron
atmospheres. From top to bottom in each panel: $\log\Tns=6.5, 6.3,
6.1, 5.9, 5.7, 5.5, 5.3$. Plotted as grey dashed lines are blackbody
spectra for the corresponding temperatures.}
\end{figure*}

\subsection{Atmosphere structure\label{s-atmostruct}}
The vertical scale of the atmosphere is transformed from the geometrical
depth $z$ to the optical depth $\tau$ by $dz=d\tau/\rho\chi$,
with $\chi$ the total mean opacity (see below) and $\rho$ the
density.  The structure of the atmosphere is then obtained by
integrating the equation of hydrostatic equilibrium
\begin{equation}
\frac{dP_g}{d\tau} = \frac{g}{\chi}
\label{e-hydrostatic}
\end{equation}
where $P_g$ is the gas pressure, from an outer boundary \taumin\ to an
inner boundary \taumax. Considering the extremely high surface gravity
encountered in a neutron star atmosphere, we neglect the effects of
radiation pressure. We compute a starting value for the integration of
eq.\,(\ref{e-hydrostatic}), $P_g(\tau=\taumin)$, assuming that the
outer layers of the atmosphere are isothermal and that the degree of
ionization is constant. With $\chi=\mathrm{const}$,
\begin{equation}
P_g({\taumin})=\frac{g}{\chi(\taumin)}\taumin.
\label{e-pstart}
\end{equation}
$\chi$ being itself a function of $(T,P_g)$, we use $\chi(\taumin)=1$
as an estimate and solve eq.\,(\ref{e-pstart}) with a Newton-Raphson
method. From $P_g(\tau=\taumin)$ we proceed with the integration of
eq.\,(\ref{e-hydrostatic}) to \taumax.  

This integration involves at each depth the evaluation of the equation
of state (EOS) as well as the computation of the absorption and
scattering coefficients. For the EOS and the radiative opacities, we
use the OPAL data \citep{iglesias+roger96-1} from RR96, that cover
three different chemical compositions: hydrogen, solar abundances, and
iron. The opacities are tabulated as a function of the temperature
$T$, $R=\rho/T_6^3$ with $T_6$ the temperature in $10^6$\,K, and
$u=E_{\gamma}/k_BT$, with $E_{\gamma}$ the photon energy (for
additional details on the OPAL opacity tables, see RR96).  For a given
depth in the atmosphere, defining $T$ and $\rho$ and at a given energy,
the radiative absorption coefficient $\kappa_\mathrm{rad}(T,\rho,E)$
is calculated from the opacity tables using a bilinear interpolation
in $\log T$ and $\log R$ and a linear interpolation in $E$. In the
construction of the model atmospheres, the radiative absorption
coefficient $\kappa_\mathrm{rad}(T,\rho,\delta E)$ for a finite energy
interval $\delta E$ is given by the harmonic mean of $n$ individual
evaluations of $\kappa_\mathrm{rad}(T,\rho,E_n)$, with $E_n$ within
$\delta E$, and $n$ sufficiently large to sample well the resolution
of the OPAL tables. 
Thomson scattering is an important source of opacity only in
low-Z atmospheres at high temperatures and high energies. The OPAL
tables implicitly include Thomson scattering as radiative opacity.
$\kappa_\mathrm{rad}$ is, hence, the sum of true absorption plus
Thomson scattering. We will come back to this issue in
Sect.\,\ref{s-radtrans}.  Finally, we compute a radiative
Rosseland mean opacity $\overline{\kappa}_\mathrm{Ross}$ from the energy
dependent  $\kappa_\mathrm{rad}(E)$.

Following RR96, we include a conductive opacity
$\kappa_\mathrm{con}=(2.5\times10^4/n_e)\times(Z^2/A)(T_7^{1/2}/\rho)\mathrm{cm^2g^{-1}}$
\citep{cox+giuli68-1} to account for energy transport by electron
conduction, with $n_e$ the number of electrons per a.m.u., $A$ the
average atomic mass, $Z=An_e$ the average ionic charge, and $T_7$ the
temperature in $10^7$\,K.  The radiative (Rosseland) opacity and the
conductive opacity are harmonically added.  The total opacity is,
hence, given by
$\chi=(\overline{\kappa}_\mathrm{Ross}^{-1}+\kappa_\mathrm{con}^{-1})^{-1}$.

Once eq.\,(\ref{e-hydrostatic}) is integrated to $\taumax$, a complete
macroscopic description of the atmosphere structure is at hand.

\subsection{Radiation transfer\label{s-radtrans}}
For the assumption given above, the radiation transfer equation takes
the form
\begin{equation}
\mu\frac{dI_{\nu}(\mu)}{d\tau_{\nu}} = I_{\nu}(\mu) - S_{\nu}, 
\label{e-radtrans}
\end{equation}
with $\mu=\cos{\theta}$ the cosine of the angle between the normal to
the atmosphere and the direction of the considered radiation beam,
$\tau_{\nu}$ the frequency-dependent optical depth, $I_{\nu}(\mu)$ the
frequency- and angle-dependent specific intensity, and $S_{\nu}$ the
source function.

As mentioned in Sect.\,\ref{s-atmostruct}, Thomson scattering is
implicitely included in the OPAL opacity tables. Considering that the
contribution of Thomson scattering to the total opacity is negligible
in high-Z atmospheres, we restrict an explicit treatment of Thomson
scattering to our hydrogen models (Sect.\,\ref{s-model_grids}). We
compute Thomson scattering, $\sigma_e=N_e\,6.65\times10^{-25}/\rho$,
but taking into account the cross-section reduction by collective
effects \citep{boercker87-1}, with $N_e$ the electron density in
$\mathrm{cm^{-3}}$, calculated under the assumption of full
ionisazion, and the true frequency-dependent absorption
$\kappa_\nu=\kappa_\mathrm{rad}(\nu)-\sigma_e$, where
$\kappa_\mathrm{rad}(\nu)$ is the radiative frequency-dependent OPAL
opacity from Sect.\,\ref{s-radtrans}.

With the isotropic and, to first order approximation,
coherent Thomson scattering term, the source function is given by
\begin{equation}
S_{\nu} = \frac{\kappa_{\nu}}{\kappa_{\nu}+\sigma_{\mathrm{e}}} B_{\nu} 
+ \frac{\sigma_{\mathrm{e}}}{\kappa_{\nu}+\sigma_{\mathrm{e}}} J_{\nu}
\end{equation}
with $B_{\nu}$ the Planck function and $J_{\nu}$ the mean intensity.
We use as boundary conditions for eq.\,(\ref{e-radtrans}) $I(\mu<0)=0$
at \taumin\ and the diffusion approximation at \taumax.
Equation\,(\ref{e-radtrans}) is solved using the well-documented
Rybicki method \citep[e.g.][]{mihalas78-1}.

For the high-Z composition atmospheres (solar abundances and iron), we
treat Thomson scattering as true absorption, directly using the OPAL
$\kappa_\mathrm{rad}$.

From the resulting angle-dependent specific intensities, the flux at
the surface of the atmosphere is computed, 
\begin{equation}
\left.
F_{\nu} = \mint{0}{1}\mu I_{\nu}(\mu)d\mu~\right|_{\D\tau=\taumin}.
\end{equation}
Radiative and conductive equilibrium requires
\begin{equation}
\mint{0}{\infty}F_{\nu}d\nu + F_\mathrm{con}= \frac{\sigma\Tns^4}{\pi}
\label{e-radequ}
\end{equation}
at each depth in the atmosphere, where $\sigma$ is the
Stefan-Boltzmann constant, $F_\mathrm{cond}=(16\sigma
T^3)/(3\rho\kappa_\mathrm{con})dT/dz$ is the conductive energy flux,
as detailed in RR96, and \Tns\ is the effective temperature of the
neutron star. If the deviations from the equilibrium (\ref{e-radequ})
are larger than a given error, correction terms $\Delta T(\tau)$ are
computed from linearization of the angle-integrated form of
eq.\,(\ref{e-radtrans})
\citep[e.g.][]{feautrier64-1,auer+mihalas69-1,auer+mihalas70-1,gustafsson71-1}.

\subsection{Model atmosphere computation\label{s-atmobuild}}
The actual computation of a model atmosphere and the emergent spectrum
is done in two separate
stages. 

(i) Starting from the temperature structure of a grey atmosphere
\citep{chandrasekhar44-1}, we iteratively construct 
(Sect.\,\ref{s-atmostruct} and Sect.\,\ref{s-radtrans}) an atmosphere
structure that satisfies radiative and hydrostatic equilibrium. At
this stage, we use an energy grid of 1000 logarithmic equidistant bins
covering the range $10^{-3.4}-10$\,keV. For each energy bin, 30
evaluations of the OPAL opacity tables are harmonically added for the
computation of the radiative absorption coefficient.  The atmosphere
structure is constructed on an optical depth grid with 100 points
covering $\tau_R=10^{-6}-5\times10^{3}$ (except for the coldest
models, which were computed on a grid covering
$\tau_R=10^{-6}-10^{3}$ due to limitations in the available
OPAL data). The angular dependency of the radiation field is sampled
by three Gaussian points in $\mu$. Radiative equilibrium
(eq.\,\ref{e-radequ} is satisfied to better then $10^{-5}$ at each
optical depth) is reached after $\sim5-15$ iterations. Generally, the
hydrogen atmospheres converged the slowest, as the steep drop of the
opacity $\propto\nu^{-3}$ makes the atmosphere transparent for
high-energy photons, thus radiatively coupling the deep hot layers
with the surface layers.
Electron conduction carries a few percent of the total flux at
the largest optical depths (Rosseland optical depths
$\tau_\mathrm{Ross}>100$) in the coldest iron and solar atmospheres.
For $\log(\Tns)>5.75$, the conductive flux is less than 1\%, which is
in good agreement with the estimates of ZPS96. Including the
conductive opacity in our atmosphere models has, hence, no noticeable
effect on the emergent spectra.

(ii) From a given atmosphere structure, we recompute the emergent
spectrum solving the radiation transfer (Sect\,\ref{s-radtrans})
on a much finer energy grid (10\,000 bins, 10 OPAL evaluations per
bin), and using 6 Gaussian points in $\mu$ to resolve the angular
dependence of the specific intensity. The higher number of angular
points has practically no influence on the angle-averaged flux from
the neutron star surface, but may be used to account for limb
darkening in the computation of spectra from neutron stars with
non-homogeneous temperature distributions.

\begin{figure}
\includegraphics[angle=270,width=8.8cm]{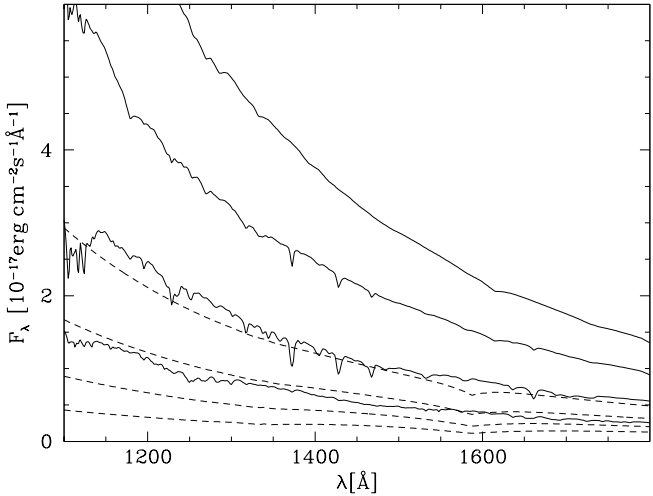}
\caption[]{\label{f-stisrange} Ultraviolet model spectra of non-magnetic
neutron stars in the wavelength range observable with HST/STIS,
redshifted for $z=0.306$. Solid lines: iron composition. Dashed lines:
hydrogen composition. From top to bottom: $\log\Tns[K]=5.7,
5.5, 5.3, 5.1$. The absolute fluxes are given for $\Rns=10$\,km,
$d=100$\,pc, and $E(B-V)=0$.}
\end{figure}

\subsection{The model spectra}
We computed grids of model atmospheres for the three different
compositions covered by the OPAL opacity/EOS tables of RR96: hydrogen,
solar abundances, and iron. We restricted ourselves to the canonical
neutron star configuration, $\Mns=1.4$\,\Msun\ and $\Rns=10$\,km,
corresponding to a surface gravitational acceleration of $\log
g=14.386$. The three grids cover the temperature range
$\log\Tns=5.1-6.5$ in steps of 0.05, i.e. a total of 29 spectra per
abundance grid.\footnote{The angle-averaged fluxes are available from
CDS, for the angle-dependent specific intensities, or models covering
additional temperatures/surface gravities, please contact the
authors.}

Figure\,\ref{f-spectra} shows the emergent spectra for the three
different compositions. As already evident in the previous neutron
star atmosphere calculations (R87, RR96, ZPS96), the model spectra
differ significantly from blackbody distributions. The pure hydrogen
models show a strong flux excess over the blackbody distributions at
energies above the peak flux. The free-free and bound-free opacity in
these atmospheres drops off rapidly at high energies, leaving Thomson
scattering as the dominant interaction between the atmosphere matter
and the radiation field. As a consequence, the atmosphere is highly
transparent to the hard X-ray photons from deep hot layers. In
contrast to the hydrogen models, the iron and solar abundance models
are overall closer to the blackbody distributions because of the
milder energy dependence of their opacities. They show, however, a
substantial amount of absorption structures, lines and edges, which
are especially pronounced in the $\sim0.1-10$\,keV range well
observable with most of the present and past X-ray satellites.

Noticeable absorption structures with equivalent widths of $\la$\,\AA\
are present in the iron spectra of moderately cool neutron stars also
in the ultraviolet.  The (non-magnetic) hydrogen models contain only
the extremely pressure broadened \La\ line. In a magnetic field \La\
splits in three Zeeman-components, with the $\sigma^{+/-}$ components
asymmetrically shifted by $\sim100$\,\AA\ for
$B\approx10^9$\,G \citep{ruderetal94-1}. Because the hydrogen in these
hot atmospheres is ionised to a large extent, the \La\ Zeeman
components are expected to be rather weak, but they may possibly be
detected with future large aperture ultraviolet telecopes, allowing a
direct measurement of the magnetic field strength.
Figure\,\ref{f-stisrange} compares the redshifted\footnote{
$z=(1-2G\Mns/\Rns c^2)^{-1/2}-1=0.306$ for an assumed $\Rns=10$\,km
and $\Mns=1.4\,\Msun$.} iron and hydrogen spectra in the far
ultraviolet.

\begin{figure}
\includegraphics[angle=270,width=8.8cm]{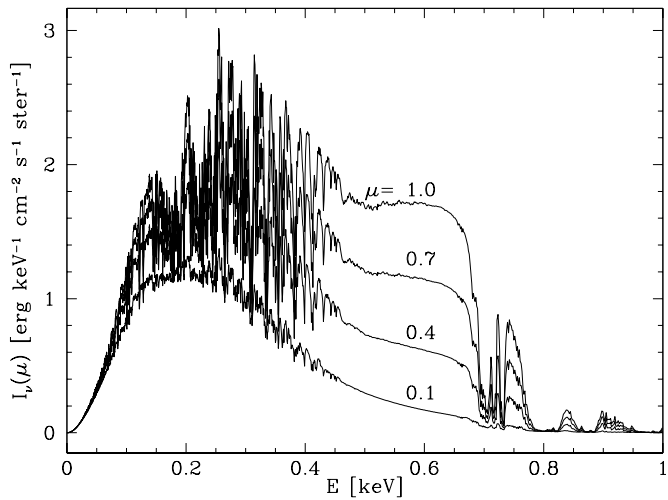}
\caption[]{\label{f-imufe} The angle-dependent specific intensity 
$I_{\nu}(\mu)$ of a $\Tns=10^6$\,K iron model for $\mu=1.0$, 0.7, 0.4, and 0.1.}
\end{figure}

Figure\,\ref{f-imufe} shows the angle-dependent intensity
$I_{\nu}(\mu)$ for a $\Tns=10^6$\,K iron atmosphere. It is apparent
that the emission from the neutron star surface is highly anisotropic,
as previously discussed by ZPS96.  Calculations of the emission of a
neutron star with a non-homogeneous surface temperature distribution
must take this anisotropy into consideration. An application of the
angle-dependent intensities will be discussed elsewhere.

\subsection{Comparison to earlier neutron star atmosphere models}
\subsubsection{\label{s-comp_rr}\citet{rajagopal+romani96-1}}
The new  model spectra agree quite well with the
models of RR96. The only systematic differences are found for the
hydrogen models, where the new models have a significantly lower flux
at energies $\ga1$\,keV.
This difference is due to the inclusion of Thomson scattering in the
new models, which increases the opacities at large optical depths in
the hydrogen atmospheres. For the heavy element atmospheres, this
effect is negligible, as the opacity is dominated by bound-bound and
bound-free absorption.

Small (5-15\,\%) differences in the emergent flux  of the heavy element
atmospheres are found in the low-temperature ($\log\Tns\le5.75$)
models in the energy range 1--10\,keV. The specific intensities at
these high energies are $15-25$ orders of magnitude below the peak
intensities, and we believe that the differences between the new
models and those of RR96 are due to the different numerical treatment
of the radiation transfer. At higher energies, the agreement between
the two different model generations is better than 1\%, except in
strong lines where different energy sampling naturally leads to
somewhat larger discrepancies.

\subsubsection{\label{s-compzvs}\citet{zavlinetal96-1}}
ZPS96 computed grids of non-magnetic neutron star LTE model
atmospheres for three different compositions, hydrogen, helium, and
iron. The hydrogen and helium opacities and non-ideal EOS employed in
these models were calculated by ZPS96 using the occupation probability
formalism described by \citet{hummer+mihalas88-1}. The iron models
were computed using the same OPAL data used also by RR96 and in the
present paper. 
The iron models show a significant difference both in the overall
shape of the continuum as well as in the details of the absorption
features.  We traced this to an erroneous use of the opacity grid in
ZPS96.  This has been confirmed (Pavlov, priv. comm.)\footnote{
After we alerted G. Pavlov to their errors in employing the opacities,
their group has computed revised models in reasonable agreement with
other work. An example revised spectrum appears in
\citep{pavlov+zavlin00-1}.}.
The ZPS96 hydrogen model is somewhat harder than both our new
spectrum and that of RR96. The flux excess of the ZPS96 model (or the
flux deficiency of our model) is not fully understood, but likely
stems from small differences in the absorption/scattering opacities.

\begin{figure}
\includegraphics[angle=270,width=8.8cm]{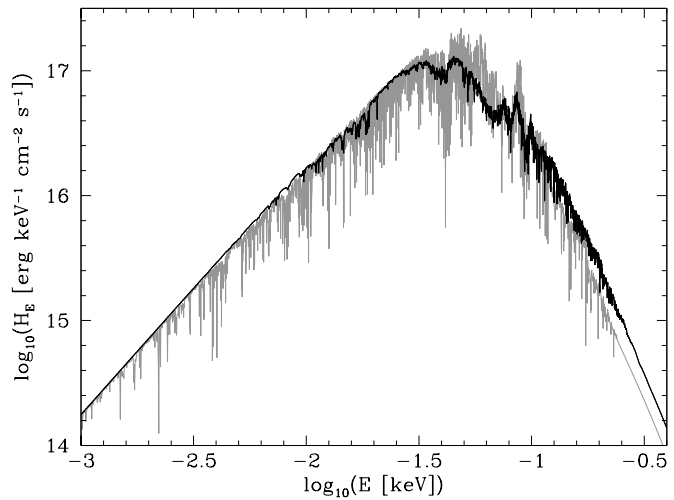}
\caption[]{\label{f-comp_werner} Comparison of the $\Tns=200\,000$\,K
pure iron NLTE model spectrum from \citet{werner+deetjen00-1} (gray
curve, unredshifted Eddington flux) with a corresponding LTE model
spectrum from our new atmosphere grid (curve).}
\end{figure}

\subsubsection{\citet{werner+deetjen00-1}}
\citet{werner+deetjen00-1} presented the first NLTE calculation for a
neutron star atmosphere using bound-free opacities from the Opacity
Project \citep{seatonetal94-1} and explicitly treated
line blanketing by millions of lines from the \citet{kurucz+bell95-1}
line list. Figure\,\ref{f-comp_werner} compares the $\Tns=200\,000$\,K
pure iron NLTE model from \citet{werner+deetjen00-1} with a
corresponding model from our LTE calculations. While the NLTE model
clearly shows more detail because of the much larger number of
considered transitions, the overall agreement between the two models
is very good. \citet{werner+deetjen00-1} quote a flux difference
between their LTE and NLTE models of $\sim10$\,\% in the line cores,
and much less in the continuum. For higher temperatures, the NLTE
effects might be stronger, but so far no detailed atomic data are
available.

\section{Beyond ``classic'' stellar atmospheres\label{s-heating}}
Part of the ongoing effort in computing neutron star atmospheres has
been carried out for the analysis of X-ray observations of isolated
neutron stars (e.g. \citealt{ponsetal02-1}). However, recent
high-resolution X-ray spectroscopy of the brightest isolated neutron
star candidates RX\,J1856.5--3754 \citep{burwitzetal01-2} apparently
excludes all the neutron star atmosphere models presented so far:
while the overall spectral energy distribution (SED) rules out 
low-$Z$ atmospheres, the featureless \textit{XMM} and \textit{Chandra}
spectra are inconsistent with the strong structures expected from a
heavy element atmosphere.

As in all previous calculations \citep[R87, RR96,
ZPS96,][]{werner+deetjen00-1}, the model atmospheres presented here
assume radiative equilibrium.  It is worth remembering that when the model
fits suggest that the flux is dominated by hot polar caps with a small
fraction of the full neutron star surface area, some sort of local
heating is likely implicated. For instance, anisotropic interior
conductivities can produce smooth variations in the surface effective
temperture; e.g. \citet{heyl+hernquist98-2}. For slow, long $P$
neutron stars, local heating can be caused by low ${\dot M}$
accretion. \citet{zampierietal95-1} and \citet{zaneetal00-1} have
produced emergent spectra for ionized H stars accreting at low rates.
With proton stopping depths of $\sim 20$\,\gc, much of the energy is
deposited fairly deep in the atmosphere, but shocks in the accretion
flow can apparently heat the outer layers, providing excess flux on
the Rayleigh-Jeans tail.
For an active magnetosphere, local heating without accretion arises
from precipitating $e^\pm$ \citep{arons81-1} or from illumination by
energetic (polar cap or outer gap) photons.  

\begin{figure}
\includegraphics[angle=270,width=8.8cm]{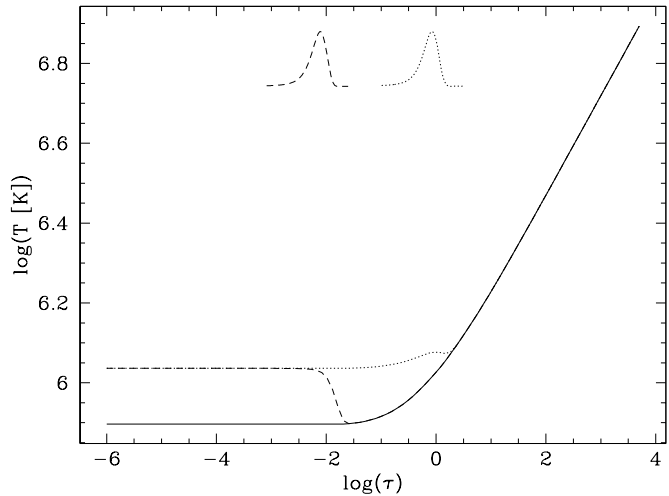}
\caption[]{\label{f-heatedatmos} Temperature structures $T(\tau)$ in
a locally heated neutron star atmosphere.  An energy amount of
$0.035\sigma\Tns^4$ ($0.82\sigma\Tns^4$) is deposited at an optical
depth $\tau_\mathrm{dep}=0.01$ ($\tau_\mathrm{dep}=1$). This energy
injection is spread by a Gaussian distribution with a width of
$\log\tau=0.3$, as illustrated in the figure, avoiding unrealistic (and
numerically problematic) step functions in the $T(\tau)$ structure.
The undisturbed temperature run is plotted as solid line, the
atmosphere heated at $\tau=0.01$ ($\tau=1$) as dashed (dot-dashed) line.
}
\end{figure}

An analogous situation --~pole caps heated by accretion~-- is
well-documented in polars, a subtype of cataclysmic variables in which
a strongly magnetic white dwarf accretes from a late type donor
star. In theses systems, the white dwarf atmosphere near the magnetic
poles is heated by strong irradiation with cyclotron radiation and
thermal bremsstrahlung from a stand-off shock. As a consequence of
this accretion heating, the spectra of these heated pole caps are
almost devoid of spectral structures. The Lyman lines, typically very
strong absorption features in the ultraviolet spectra of white dwarfs,
are almost completely flattened out \citep{gaensickeetal98-2}, and
also the X-ray spectra of polars are very close to blackbody
distributions,  unlike the expected emission from a high-gravity
photosphere strongly enriched with heavy elements
\citep{mauche99-2}. While the physical processes involved in heating
the pole caps of accreting white dwarfs and accreting neutron stars
differ markedly, the observed phenomenology is very similar:
RX\,J0720.4--3125, another bright isolated neutron star shows a
quasi-sinusoidal X-ray light curve suggesting the presence of large
heated pole caps, whereas the X-ray spectrum contains no significant
absorption structure \citep{haberletal97-1,paerelsetal01-1}
reminiscent of the ultraviolet observations of the heated white dwarf
in the polar AM\,Her \citep{gaensickeetal98-2}. It appears likely, therefore,
that heating effects may affect the line spectra of heavy
element neutron star atmospheres.

\begin{figure}
\includegraphics[width=8.8cm]{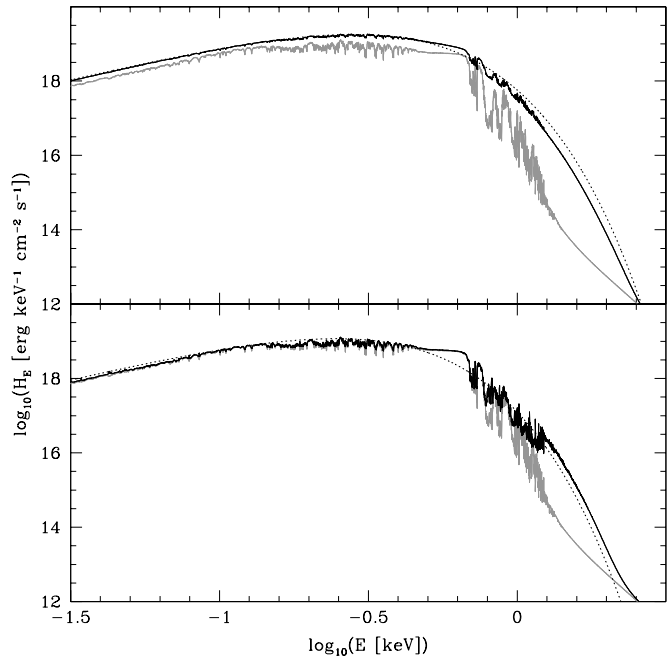}
\caption[]{\label{f-heatedspectra} Emergent spectra from a heated
$\Tns=10^6$\,K iron atmosphere. Top panel: $0.82\sigma\Tns^4$ are
deposited at $\tau_\mathrm{dep}=1$, equivalent
$\Sigma=3.2\times10^{-3}$\,\gc. The spectrum of the heated atmosphere
and of the undisturbed atmosphere are plotted as black and as gray
lines, respectively. The dashed line shows a blackbody spectrum
corresponding to the effective temperature of the heated
atmosphere. Bottom panel: $0.035\sigma\Tns^4$ are deposited at
deposited at $\tau_\mathrm{dep}=0.01$, equivalent
$\Sigma=2.2\times10^{-4}$\,\gc.}
\end{figure}

As a simple illustration, we ignored the details of particular
physical heating effects and computed a number of spectra under the
assumption that an energy $\alpha\sigma\Tns^4$ is deposited at a
characteristic depth in the atmosphere, $\tau_\mathrm{dep}$
(Fig.\,\ref{f-heatedatmos}). In practice, the emergent spectrum of a
``heated'' neutron star is computed by (1) modifying the thermal
structure of an undisturbed atmosphere $T(\tau)$ to account for the
deposited energy, (2) recomputing the atmosphere structure based on
the modified temperature profile (Sect.\,\ref{s-atmostruct}), and
finally solving the radiation tansfer within the ``heated'' atmosphere
(Sect.\,\ref{s-radtrans}). 
Figure\,\ref{f-heatedspectra} shows the emergent fluxes from a
$\Tns=10^6$\,K iron atmosphere for two different sets of parameters,
$\alpha=0.82$ and $\tau_\mathrm{dep}=1$, equivalent to
$\Sigma=3.2\times10^{-3}$\,\gc, and $\alpha=0.035$ and
$\tau_\mathrm{dep}=0.01$, equivalent to
$\Sigma=2.2\times10^{-4}$\,\gc.

Two effects are evident. The flatter temperature in the outer layers
of the heated atmospheres supresses the strong absorption edge around
1\,keV, returning the Wien tail colour quite close to that of a simple
blackbody. Further, the equivalent widths of the lines are, of course,
strongly affected. Heating at $\tau \sim 1$ strongly supresses the
$\le $\,keV absorption features; surface heating in fact drives the
line features on the Rayleigh-Jeans tail into emission.

We stress that this illustrative approach is by no means a
self-consistent model of a heated neutron star atmosphere, as we 
include neither any physical assumption on the actual heating
mechanims, e.g., there are currently no plausbile physical processes
known which are capable of heating such shallow layers, nor a proper
energy balance of the heating/cooling processes. It is merely intended
as a possible explanation for the apparent paradox that the SEDs of
isolated neutron stars (in particular RX\,J1856.5--3754) are
best-fitted with heavy-element model spectra
(\citealt{pavlovetal96-1}; this paper; \citealt{burwitzetal01-2};
\citealt{ponsetal02-1}), whereas their X-ray spectra show at best very
weak spectral structures.

More detailed future work will need to carefully examine the
efficiency of the various possible heating mechanisms, in particular
to establish the depth-dependent energy deposition, and to account
properly for the energy balance in the neutron star atmosphere.

\section{Conclusions}
We have presented a set of angle-dependant low field neutron star
atmosphere spectra. These radiative and hydrostatic equilibrium
atmospheres should be useful to researchers pursuing exploratory fits
of soft X-ray/UV/optical data. However, it is important to inject a
note of caution into the discussion. At a minimum, non-uniform surface
temperature, combined with limb-darkening and gravitational focussing
will have subtle, but important effects which can be modeled using the
present grid.

Many other physical effects can, of course, strongly perturb neutron
star spectra. The effects on heavy element absorption lines and edges
can be particularly strong.  Magnetic fields characteristic of young
radio pulsars will have a dramatic effect on the opacities and the
emergent spectra. When considering the strength of the absorption
lines in spectra from magnetic atmospheres with heavy elements
\citep{rajagopaletal97-1} it is important to remember that the line
energies depend sensitively on the B-field. Even simple dipole $B$
variation across a polar cap can shift line energies by $\sim 10$\%,
strongly decreasing the equivalent width of spectral features in
phase-averaged spectra. Convection has also been considered
(eg. RR96), although even small magnetic fields may supress convective
transport and keep atmospheres close to the hydrostatic solution.

On the whole, while we feel that our atmosphere grid should be of use for
fitting of observed data; complications  are certainly
expected.  In particular strong $B$-field variations and surface
re-heating can decrease the equivalent width of heavy element line
features in neutron stars with active magnetosphers; this may explain
in part the difficulty in finding such features in early {\it Chandra/XMM}
data \citep[e.g.][]{paerelsetal01-1, burwitzetal01-2}.

\acknowledgements{BTG was supported in part by a travel grant of the
Deutscher Akademischer Auslandsdienst (PKZ: D/99/08935) and by the DLR under
grant 50\,OR\,99\,03\,6. We thank K. Werner and G. Pavlov for
providing their model spectra for a quantitative comparison, and
B. Rutledge for comments on an earlier draft.  We thank the referee,
Dr. Zavlin, for useful comments.}

\bibliographystyle{aa}


\end{document}